\documentclass[11pt,twoside]{article}

\usepackage{asp2006}
\usepackage{epsf}
\usepackage{lscape}

\begin{document}
\title{Primordial light element abundances}   
\author{Paolo Molaro}   
\affil{ INAF Osservatorio Astronomico di Trieste, I-34143 Trieste, Italy}    

\begin{abstract} 

After few minutes the Universe   evolved through conditions of temperature and density which
permitted the first synthesis of astrophysically interesting abundances of D, $^3$He,
$^4$He and $^7$Li. The relic abundances   are sensitive  probes of the nucleon density and  so are the 
   CMB acoustic oscillations, somewhat 400000 years later, which  allow a stringent cross check.
The CMB high precision estimate of the   baryon density  by WMAP  is currently 
used as input parameter for  standard big bang nucleosynthesis  (SBBN) to interpret primordial abundances
rather then  being directly derived
from the observations of light elements as  was common   use before. New atomic physics and identification of systematics 
lead to an upwards revision of the 
$^4$He primordial abundance at Yp=0.2477$\pm $0.0029  (Peimbert et al 2007) 
 removing a major source of tension between  SBBN and  WMAP. 
 The D/H as measured in QSO high redshift absorbing clouds  shows an excess of scatter but  the mean value  is 
 found in  spectacular agreement with the WMAP-  $\Omega_b$ prediction.
 The  Li/H  recently   redetermined     in halo dwarfs is 
more than a factor 4 lower than expected. We argue that the difference reduces to a  factor 2  when the IRFM  Teff 
scale is adopted. Diffusion has been  suggested to have depleted Li in halo dwarfs by the required amount to remove  the  gap, however this would imply an implausible high abundance   of the 
more  fragile $^6$Li detected in some 
halo dwarfs, thus    leaving the puzzle open.
\end{abstract}


\section{John Beckman and the light elements}   
The organizers  asked me to  review  the status of the light elements, which is one of  J.E. Beckman's interests although  a bit aside the main 
theme of this unique and enjoyable  conference.
John's  interest on light elements dates back to  the 80's  at the epoch of his definitive move to  IAC.
The  start was a search for BeII lines in HD76932 by means of IUE spectra, a star  belonging to the first sample of halo stars
were Francois  and Monique Spite observed the primordial  Li. The    upper limit  provided for Be  showed that  spallation processes 
of high energy cosmic rays were 
not an alternative to primordial nucleosynthesis to make up for  all
 or a fraction
of the Li observed in the star, thus supporting its primordial origin (Molaro \& Beckman 1984). This  work was followed by the first Be detections in halo stars  by means of the Image Photon Counting System at the Isaac Newton Telescope
on La Palma  
and reported in Rebolo et al (1988a). In the same years a Li observational campaign was conducted from La Palma with  INT and first results presented  at the second IAP workshop 
in Paris  (Beckman et al 1986),  and then in a more  definitive shape in  Rebolo et al (1988b).
This was the first confirmation of the primordial  Spite and Spite  halo plateau made by and independent group, significantly enlarging the 
sample of halo stars in number and physical properties, in particular   with a significant extension in metallicity down to [Fe/H]=-3.5, which is not very far from where we are today.
   After these pioneering works the research of light elements became very active  at the IAC where  R. G. Lopez and 
E. Martin, among the others, extended the work to other light elements  (B)  and to other environments such as 
stellar clusters, cool x-ray-binary companions  and brown dwarfs. A field  still very active today at IAC.

 \section{The WMAP-$\Omega_b h^2$  primordial abundances}

 The number of baryons  is set once for
 all at the baryogenesis and is constant in a comoving universe.  When the universe was few minutes old  
 it    evolved through conditions 
 of temperature and density which
permitted the first synthesis of astrophysically interesting abundances of D, $^3$He,
$^4$He and $^7$Li. 
 While the $^4$He abundance is determined primarily
 by the universal expansion rate and  is sensitive to the number of relativistic
 particles at the nucleosynthesis epoch,  D and Li abundances depend strongly
 from the available nuclei.   In the standard BB nucleosynthesis with 3 neutrino flavours the only free
 parameter is  the relative number of  baryons to photons, namely  $\eta = n_{b}/n_{\gamma}$ .
 In the  range  of interest 3$<$$10^{10} \eta < 8 $,  D/H and Li/H  show the strongest dependence on 
 $ \eta$  and therefore on the particle density, these are   $\propto \eta ^{-1/6}$ and $\propto \eta ^{2}$, respectively.

 The value of the baryon density can be measured from the acoustic oscillations
 in the CMB since the baryonic component determines the inertia of the
 mass-photon fluid. The ratio of the compressible peaks of CMB power spectrum are sensitive to the
 baryonic density and allow a very precise measure of $\Omega _{b}$ .
 Considering the WMAP-Only  3 year data, 
 Spergel et al (2007) derived  $\Omega_b h^2$ =0.02233 $^{+0.00072}_{-0.00091}$.

 The number of photons is 
 also constant  in a comoving universe after  e$^\pm$
 annihilation and   therefore   
 $\eta = 10^{10} (n_{b}/n_{\gamma})_0 = (273.9 \pm 0.3) \Omega _{b}h^2$, where the small error enters in  the conversion from number to  mass  
 densities (Steigman 2006).  The WMAP  value for $\Omega_b h^2$  yields 
  $\eta = 10^{10} (n_{b}/n_{\gamma})_0 = 6.116 ^{+0.197}_{-0.249}$ which
  fixes  the primordial abundances of  the
 light elements in the framework of SBBN.   By  using  Kneller and Steigman (2004) approximations of primordial yields in the 
 range  3$<$$10^{10} \eta < 8 $  we obtain the predicted primordial abundances reported    in Table 1, where the primordial $^4$He, Y$_p$, is given in mass fraction and the other light elements in number fractions with respect to hydrogen.

\begin{table}[!ht]
\label{sizes}
\caption{WMAP primordial abundances}
\smallskip
\begin{center}
{\small
\begin{tabular}{ll}
\tableline
\noalign{\smallskip}
Element &  SBBN+WMAP \\
\noalign{\smallskip}
\tableline
\noalign{\smallskip}
Yp & {0.2482 $^{+0.0004}_{-0.0003}$}\\
$^3He/H$ &{(10.5$\pm 0.6)$ $\cdot 10^{-6}$}\\
D/H&{(25.7 $^{+1.7}_{-1.3}) \cdot 10^{-6}$}\\
Li/H&{ $(4.41 ^{+0.3}_{-0.4}) \cdot 10^{-10}$}\\
\tableline
\end{tabular}
}
\end{center}
\end{table}

However, some
  degeneracies  are present in the derivation of the cosmological parameters from the CMB power spectrum. In particular  $\Omega_b h^2$ depends from the index of the power law of primordial fluctuations and from the opacity
 at the reionization epochs. For instance, the WMAP 3 year combined with all other relevant datasets   with a running spectral index is 
 providing $\Omega_b h^2$ =0.02065 $^{+0.00083}_{-0.00082}$.   This value gives   
  $\eta _{10} = 5.656 ^{+0.227}_{-0.225}$ and the predicted primordial abundances  become: Yp= 0.2474 $^{+0.0004}_{-0.0003}$, D/H=  (29.1 $^{+1.9}_{-1.8}$) $\cdot$ 10$^{-6}$ and
 Li/H= (3.8 $\pm$ 0.4) $\cdot$ 10$^{-10}$ with    about 11\% shift upwards for D/H abundance  and 14\% downwards for Li/H. Thus, uncertainties
 regarding priors are a significant source of systematic errors which  slightly exceed the statistical error in the
 prediction of $\Omega_{b} h^2$. Nevertheless, the CMB  precision estimate of the   baryon density   remains high  and should be 
used as input parameter to interpret primordial abundances
rather then  being directly derived
from the observations of light elements as it was usual before WMAP. Only an unlikely knowledge of a primordial D or Li abundance with an accuracy greater 
 than 10-15\% could  reverse the logical path.
 
 \section{Helium}
 Primordial helium is best determined through recombination of HeII and HII
 lines in extragalactic, almost chemically unevolved HII regions, with an extrapolation to zero metallicity
 to account for stellar production.  Fig. 1 shows the behaviour of the $^4$He 
 determinations in the last
 decades.  After many years with rather low values we have now approached the
 WMAP values thus removing a major source of tension between WMAP and BBN.
 Olive and Skillman (2004) stressed the importance of  systematic errors and
 derived 
 Y$_p$ = 0.2495 $\pm 0.0092$., while  Peimbert et al  (2007)  make a list of   13 different types of 
 errors, several of systematic origin, entering in the primordial $^4$He determinations.  Fukugita and Kawasaki (2006) considered carefully the underlying $^4$He stellar absorption and 
 derived Y$_p$ = 0.250 $\pm 0.004$.
  Peimbert, Luridiana \& Peimbert (2007)   derived 
 Y$_p$ = 0.2477 $\pm 0.0029$ by using   the new
 recombination coefficients of HeI from Porter et al (2005) and excitation collisional coefficients for H from Anderson et al (2002). This new physics alone
 determine a 0.008 dex increment in the $^4$He determination. 
 By using the Porter et al emissivities Izotov et al (2007) obtained Y$_p$= 0.2516$\pm$0.0011 which exceed by 2$\sigma$ the predicted WMAP value implying a slight deviation, but of different sign, from SBBN.

  \setcounter{figure}{0}
 \begin{figure}[!h]
 \plotfiddle{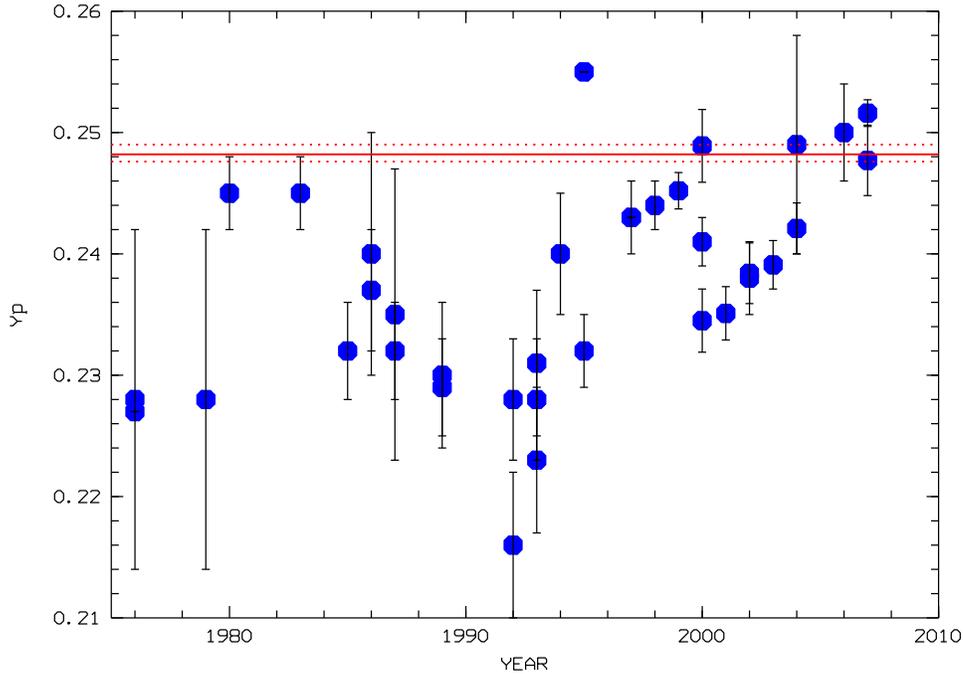}{2.5in}{-90}{50}{50}{-220}{288}
 \caption{ Historical  record of $^4$He determinations. The WMAP-SBBN  predicted primordial value is also shown with the red line and 1$\sigma$ error. }
 \end{figure}
 
 $^3He$ has a complex post-BBN evolution with unclear stellar production. Bania, Rood and Balser (2002) after 20 years of heroic observations of  the most distant Galactic HII region suggest
 11$\pm 2$ ppm (parts-per-million)  as the best upper limit on the primordial   $^3He$  abundance, essentially derived from the observation of a single HII nebula. This is in excellent agreement with the WMAP value 10.5 $\pm 0.6$ ppm, but considering the large uncertainties involved  $^3He$ is not 
 imposing strong constraints on SBBN.

 \section{Deuterium}
 
 Deuterium  
 has  a very simple  chemical evolution history. It cannot be produced by other sources
  than the BBN and whenever it is cycled in stars it is entirely burned out. 
The possibility of observing primordial D/H in distant chemically unevolved clouds was predicted already by Adams (1976)
 but the first observations were those of Tytler et al (1996) and after a decade of 8-10 m large 
 telescope observations we remain with  only   8 measurements. 
  The observations   are shown  in Fig. 2   where they are plotted  with respect to the neutral hydrogen column densities which imply  different types of absorbers.   The sub-DLA cases with hydrogen column densities LogN(HI)  $\le$ 19 cm$^{-2}$ between are those which better match the WMAP value. 
 Kirkman et al (2003) claimed for the presence of a correlation between D/H and HI, but the new measurements of O'Meara et al (2006) of 33.1 $\pm$ 4.3 ppm (parts-per-million) at LogN(HI)=20.67 cm$^{-2}$ and the one  of Crighton et al (2004)
 16.0 $\pm$ 2.5 ppm at logN(HI)=18.25  cm$^{-2}$ showed  that the supposed correlation with HI  was an artifact produced by poor statistics. The systems show low  metallicities  in the range  -3$<$ [Si/H] $<$ -1 and no hint of correlation between D/H and metallicity or redshift. At these metallicities 
no significant astration is foreseen according to Romano et al (2003) computations, and
 the measurements should reflect essentially pristine gas composition. 
  From Fig. 2  it is evident that the dispersion of the measures exceeds the reported errors suggesting at face value either the presence
 of a scatter in the D/H or an underestimation of the errors in some or all  measurements.
 The  unweighted average of the D/H gives 28.1$\pm$9 ppm, with the central value  in excellent agreement with the 25.7 $^{+1.7}_{-1.3}$ ppm of WMAP. It is 
 really a very unfortunate case
 that we do not  understand the cause  of the observed scatter  since it prevents the potential use of D/H into the 
 identification of the priors in the CMB analysis.
  
 A  matter of concern are   the Galactic D/H abundances, which turn out to be more complex than the high redshift universe.
   Fuse has provided a lot of new data confirming the dispersion in the galactic disk
 values with a bimodal distribution of high  (D/H=23.1$\pm$ 2.4  ppm) and low values (10.1$\pm$3.3 ppm) once cleaned from the local bubble values, which cluster onto  a third value (15.8$\pm$ 2.1 ppm)
 (Savage et al 2007, Linsky et al 2006).
 The measurements show a correlation with  Fe abundances suggesting  a depletion of D into dust grains. A new measurement of the deuterium hyperfine transition towards the Galactic anticenter region  of  21 $\pm$ 2.3 ppm is also consistent with the high D/H values in the solar neighborhood
 (Rogers et al 2007).
The  high values should  therefore be more representative of the gas phase abundances but they impose tight constraints on the amount of minimum recycling foreseen in the 
 Galactic chemical evolution and require   infall of primordial  D
 (Steigman et al 2007).   
 
 \setcounter{figure}{1}
 \begin{figure}[!h]
 \plotfiddle{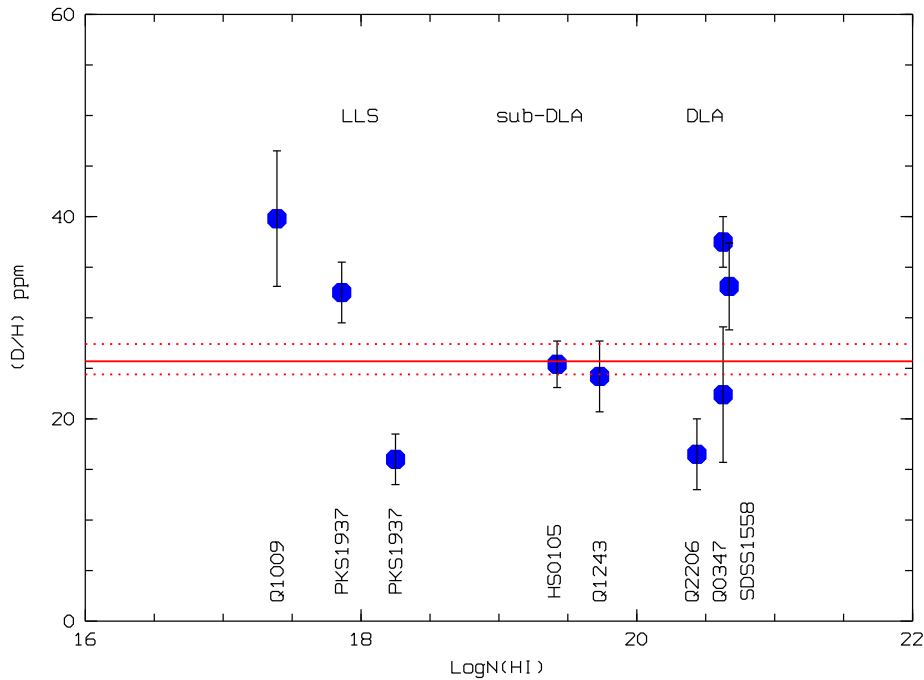}{3.5in}{-90}{50}{50}{-220}{288}
 \caption{D/H versus hydrogen column density. The lables identify the QSO where the D/H has been measured: Q1009+299 is from Tytler et al (1996), PKS1937-1009 are from  Burles and Tytler (1998) and from Crighton et al (2004), HS0105 +1619 is from O'Meara et al ( 2001); 
  Q1243+3047  is  from Kirkman et al (2003),  Q2206-199 is from  Pettini and Bowen (2001); for Q0347-383 there are two determinations of the same system by D' Odorico et al (2001) and by Levshakov et al (2002); SDSS1558-0031is from O'Meara et al (2006) }
 \end{figure}

\section{Lithium}    

\subsection{ Off, but by how much?}

The more significant Li measurements in halo stars since  Spite and Spite (1982) are shown in Fig.  3, where A(Li)=($\log (Li/H)+12)$ . At variance with the other elements there is a clear  gap
between the measured values and the WMAP predictions, which seems to increase with time instead 
of  converging  onto the WMAP value.
 The most recent works are 
Asplund et al (2006), characterized by data with very  high signal-to-noise, and  
 Bonifacio et al (2007) who  increases significantly 
the data points at low metallicity. Their results recomputed  on common scales 
 are shown in  Fig 4  taken from Bonifacio et al
(2007).   Fig. 4  shows   an increase in the dispersion at low metallicity, the presence of  a slope with metallicity,  or more precisely of a discontinuity
in the Li abundance at [Fe/H]$\approx$ -2.5 and even 
a possible drop out of  Li abundance at about [Fe/H]=-3.5. 
Both works conclude for  a rather low Li primordial value  somewhat around A(Li)= 2.1, or even 2.0  if the observed slope 
is extrapolated at metallicities of Fe/H=-3.5. Thus, we note that  the most recent values are very  close to the initial value of 2.05 
proposed   by the Spites  more than 20 years ago. 

 \setcounter{figure}{2}
 \begin{figure}
 \plotfiddle{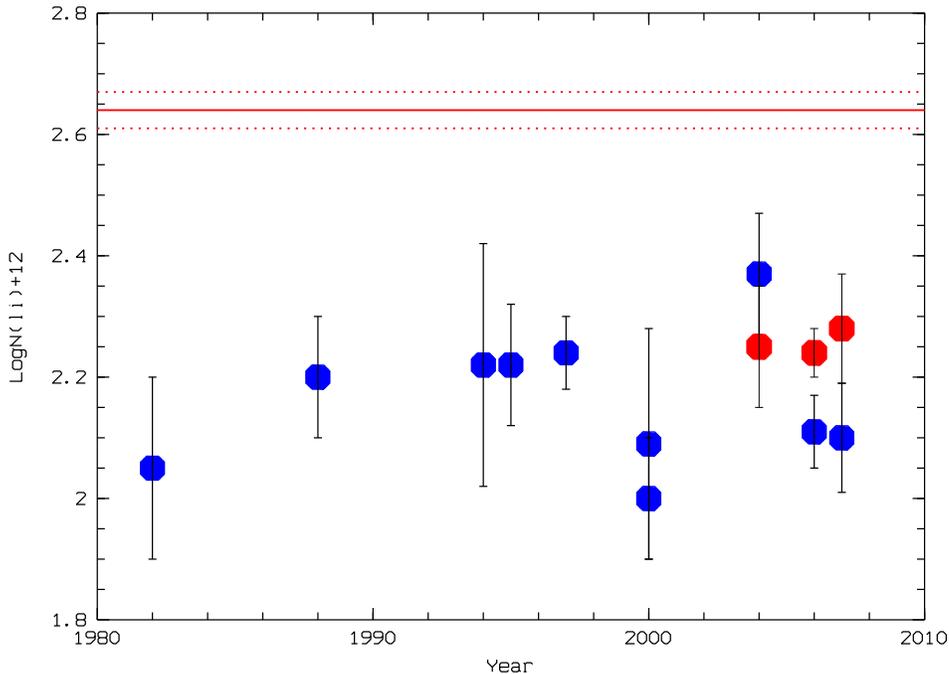}{3.5in}{-90}{50}{50}{-220}{288}
 \caption{  Filled blue circles are the historical record of primordial Li/H determinations in field halo stars since  Spite and Spite (1982). The filled red circles are re-determinations after correction for some systematics  as explained in the  text. A(Li)=($\log (Li/H)+12$) }
 \end{figure}
 
  Li is mostly ionized and remains very sensitive 
to the T$_{eff}$. 
At temperatures of interest an error of 100 K in T$_{eff}$ translates into about 0.08 dex in Li abundance. 
Both  Asplund et al (2006) and Bonifacio et al (2007)
use  T$_{eff}$  obtained through 
H$\alpha$ fitting and adopting the Barklem et al (2000, 2002) theory of collisional broadening. 
This   new treatment provides lower Li values by $\approx$ 0.1 dex at lower metallicities with comparison to
the Ali Griem  (1966) theory generally adopted in   the  preceding works. 
Thus,
 these  T$_{eff}$ are  likely responsible of the abrupt change  in the Li 
 abundances observed  at [Fe/H]$\approx$ -2.5, and of the low values obtained at lower metallicities.
  For instance, 
keeping  only the sample of 19 stars with metallicity [Fe/H]$>$ -2.5  from  Asplund et al (2006) we obtain  A(Li) =2.24 $\pm$ 0.04.
Moreover, the data of  Bonifacio et al (2007) but with different  T$_{eff}$ obtained by using the  Ali-Griem (1966) theory provide the value of A(Li) =2.28 $\pm$ 0.09, 
significantly enhancing by $\approx$  0.2 dex the average  value  (Bonifacio et al 2003).

 \setcounter{figure}{3}
 \begin{figure}
 \plotfiddle{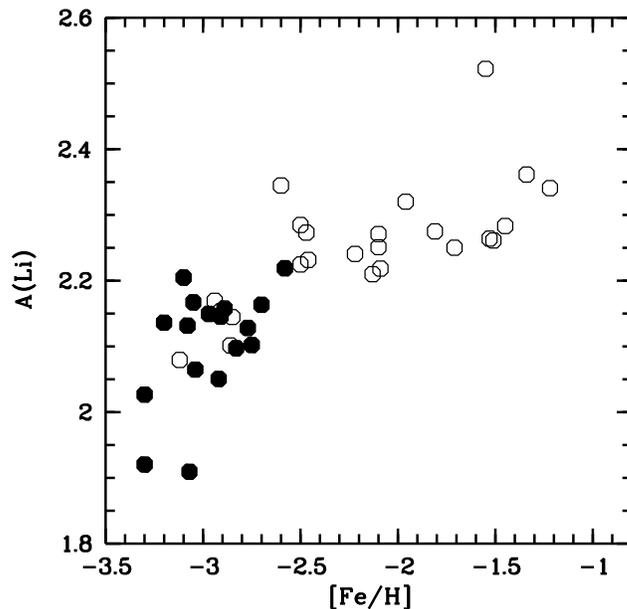}{3.5in}{0}{100}{100}{-200}{-100}
 \caption{ Li/H from Asplund et al (2006) (circles) and Bonifacio et al (2007) (filled circles) plotted on a common scale, from Bonifacio et al (2007)}
 \end{figure}
 
As pointed out in Molaro et al (1995)  the adoption of  $T_{eff}$  obtained with the 
Infra-Red-Flux-Method  is levelling out 
  any  slope of Li with  metallicity .  This  has been also confirmed by      Bonifacio and  Molaro (1997) and  Melendez and Ramirez  (2004) which make use of  the IRFM  $T_{eff}$ and, while sharing  several stars in common with Asplund et al  (2006), do not find any evidence of Li slope with metallicity. 
Fig. 5 of Asplund et al   compares the $T_{eff}$ obtained 
 from H$\alpha$ fitting plus Barklem theory for
 collisional broadening with those obtained with   IRFM  method and reveal  a  divergence  between the two scales at low metallicities, i.e. [Fe/H]$<$ -2.5. 
 The IRFM  method faces problems with reddening correction and with the difficulty of having accurate K magnitudes, but 
 is  less  model dependent, and therefore less sensitive to our ability of modeling the stellar atmosphere. Moreover, 
 Barklem (2007) points out that NLTE effects could  be important in the wings of the balmer lines thus questioning the
T$_{eff}$ obtained in this way.

Melendez and Ramirez (2004) by means of IRFM  value obtained  A(Li)=2.37 $\pm$ 0.06. However, this value  cannot be taken at face value since  it is obtained by 
means of   stellar atmospheric models which  provide  systematically higher Li abundance than other  codes.
They use the Kurucz models which 
adopt an approximate treatment for overshooting
which has been shown to provide an unsatisfactory color match at low metallicity and should not be used (Castelli et al 1997). Considering
this effect their value should be lowered by about 0.08 dex,  plus an additional correction  for the 
 NLTE correction not taken into account, to reach a final value of $\approx$  2.25.

Thus,  eliminating the data values obtained with T$_{eff}$ from Barklem theory at low metallicities, which is likely affected by some systematics, 
  it is possible to find  a convergence of the various measures and methods
 around a common range of  values between  A(Li)= 2.24-2.28. This   still  remains   significantly below
  the WMAP predicted value, but not as much as
 the factor 4  implied by the Asplund et al (2006) and Bonifacio et al (2007) measures.

The hypothetical presence of 
a drop in Li at very low metallicities may find  support also in the absence of Li  in  HE 1327-2326 a dwarf with [Fe/H]=-5.45  (Aoki et al 2006).  However, HE 1327-232  is rather chemically peculiar with 
[C/Fe]=+4 and the chemical composition may reflect single supernovae contamination were most of the Li is burned out. On the other hand  the
 more  "normal" metal poor dwarf  CS 22876-32  with  [Fe/H]=-3.7 shows  A(Li)=2.2   in the main component of the binary system
suggesting a normal behaviour of Li at very low metallicities (Gonzalez-Hernandez et
al 2007). 

\subsection{Is the Li puzzle solved?}
Stellar depletion either by rotationally induced mixing or diffusion have been suggested as possible mechanisms able to reduce the Li
abundance in halo stars.
These models predict  more dispersion in the Li abundances  of what is observed in the halo stars and a pronounced down turn in the Li abundances at the hot end of the Li plateau.
Richard et al (2005) invoke some extra turbulence to limit the diffusion in the hotter stars and to restore uniform Li abundance along all the plateau.
Recently, Korn et al (2006) claimed the detection of a diffusion ``signature''
in the Globular Cluster NGC 6397. According to their analysis the Turn Off stars
in this cluster exhibit lower iron and lithium abundances than the slightly more
evolved stars. These observations may be interpreted in the framework of the
diffusive models of Richard et al  (2005) and this has been regarded as 
the  solution of the Li problem (Charbonnel 2006).

The Korn et al  result is very suggestive; however, it again
relies heavily on the adopted temperature scale and an increase by only 100 K of the
effective temperature assigned to the TO stars would remove the abundance
differences between these  and subgiant stars. Previous analyses of the same
cluster by Castilho et al (2000) and Gratton et al (2001), with different assumptions on the effective
temperature scale, failed to find any difference in [Fe/H] between TO,
subgiant, and giant stars.
 Bonifacio et al (2002)  studying the Li in the  this cluster obtained a slightly hotter
temperature for the TO and A(Li)=2.34  which suffices to eliminate any    Li difference between the TO and subgiants.
 At  deep inspection globular clusters  show Li scatter among the mean
sequence stars which cannot be related to diffusive processes since they  show  Li-O and Li-Na anticorrelations. These have been observed   in NGC 6752  (Pasquini et al 2005) 
and 47 Tuc (Bonifacio et al 2007) and suggest that the formation of globular clusters could be a complex  process where it may be  difficult  to disentangle small  diffusive signatures.

The diffusion hypothesis  encounters an additional problem with the  $^6Li$ detection 
 in some  halo stars by Asplund et al. (2006). 
In the presence of diffusion  of the sort required to reproduce the $^7Li$  expected for WMAP,  
 $^6Li$   is depleted by the same processes  in an even larger  amount. When pre-main sequence destruction is also considered the
  resulting $^6Li$   original value is almost comparable to the $^7Li$  level. As noted by Asplund et al 92006)  the  Richard et al (2005) model  T6.25  when applied to the star 
  LP 815-43 implies an unplausible high original $^{6}$Li of $\approx$ 2.6.   
This level of $^6Li$  would pose serious problems for the $^6Li$  synthesis, so that the presence of the Li isotope challenges the diffusion
hypothesis (cfr Prantzos 2007)

The reason for the  disagreement of the observations with the WMAP value is not readily identified. 
Atmospheric modeling problems are unlikely but not completely ruled out. Observations of the Li subordinate, which forms in a different
layer in  the atmosphere, provide consistent results to that of the main Li line (Bonifacio and Molaro 1998, Asplund et al 2006).
First results from 3D atmospheric models     show 
 very small departures from 1D NLTE  Li
abundances  with small but negative  corrections thus eventually exacerbating the disagreement with
 the WMAP based prediction (Asplund et al 2003, Barklem et al 2003).

A nuclear fix to the Li problem is also unlikely (Cyburt et al 2004). 
Coc et al (2004)   suggested that the  cross section of   $\rm ^7Be$ $(d,p)$ $\rm 2^4He$ and $^7$Be$(d,\alpha)^5$Li could be larger, but recent 
measures of  this cross section at energies appropriate
to the big bang environment  found it  10 {\em smaller}, than
leaving not  much room for a Li change due to nuclear cross sections (Angulo et al 2005).

Piau  et al (2006) suggested that significant  matter  processing  in  pregalactic PopIII stars could lower the Li abundance and also explain 
 the presence of Li dispersion  
among the metal poor sample as well as the fact the absence of Li in HE 1327-2326. However, as pointed out by Prantzos (2007) the considerable amount of matter recycling in PopIII
stars should  produce   significant amounts of heavy   elements, which are not seen.

Although a solution of the Li problem  may still reside in our ability to reconstruct the complexity of the stellar atmosphere, systematic errors in the
T$_{eff}$ or 3D effects   able to increase  the Li value to match the WMAP value  seem  unlikely.  It is thus possible that
  the disagreement  may be  real revealing  the presence of  unaccounted processes  in the BBN. 
 Neutral  decaying particles   (Jedamzik  2006, but see Ellis et al 2005) or  a variation of the deuteron binding energy (Dent et al 2007) have been already considered and 
 are among the more intriguing  scenarios  for the solution of the  puzzle.


\acknowledgements 
It is a pleasure to thank the organizers for their kind invitation to  this unique  conference and to
Piercarlo Bonifacio, Valentina Lauridiana and Bernard Pagel for enjoyable discussions on the light elements. Special thanks  to John Beckman for his advice and inspiration. 


\end{document}